\documentclass[12pt]{article}
\usepackage{times}
\usepackage{geometry}
\usepackage[utf8]{inputenc}
\usepackage{enumitem,amssymb}
\usepackage{ragged2e}
\usepackage[font=footnotesize]{caption}
\newlist{thematic}{itemize}{8}
\setlist[thematic]{label=$\square$}
\usepackage{pifont}

\newcommand{\Xmax}{\ensuremath{X_{\rm max}}}

\usepackage{hyperref}
\usepackage{epsfig}
\usepackage{color}
\usepackage{bm}
\usepackage{mathrsfs} 
\usepackage{amsmath}
\usepackage[numbers,sort&compress,square]{natbib}
\usepackage{wrapfig}
\usepackage{url}
\usepackage{titlesec}
\usepackage{pdfpages}

\titlespacing*{\section}{0pt}{0.2\baselineskip}{\baselineskip}

\usepackage[usenames,dvipsnames]{xcolor}
\definecolor{orange}{cmyk}{0,0.5,1,0}
\definecolor{rossoCP3}{cmyk}{0,.88,.77,.40}
\definecolor{graa}{rgb}{0.8,0.8,0.8}
\definecolor{blaa}{rgb}{0.2,0.2,0.6}
\definecolor{darkBlue}{rgb}{0, 0, 0.8}
\definecolor{darkGray}{rgb}{0.3, 0.3, 0.3}

\hypersetup{
    bookmarksopen=true,     
    bookmarksopenlevel=1,
    unicode=false,          
    pdftoolbar=true,        
    pdfmenubar=true,        
    colorlinks=true,        
    linkcolor=darkBlue,     
    citecolor=darkBlue,      
    filecolor=darkBlue,      
    urlcolor=darkBlue        
}

\usepackage{lineno}

\begin{document}

\begin{titlepage}
\includepdf{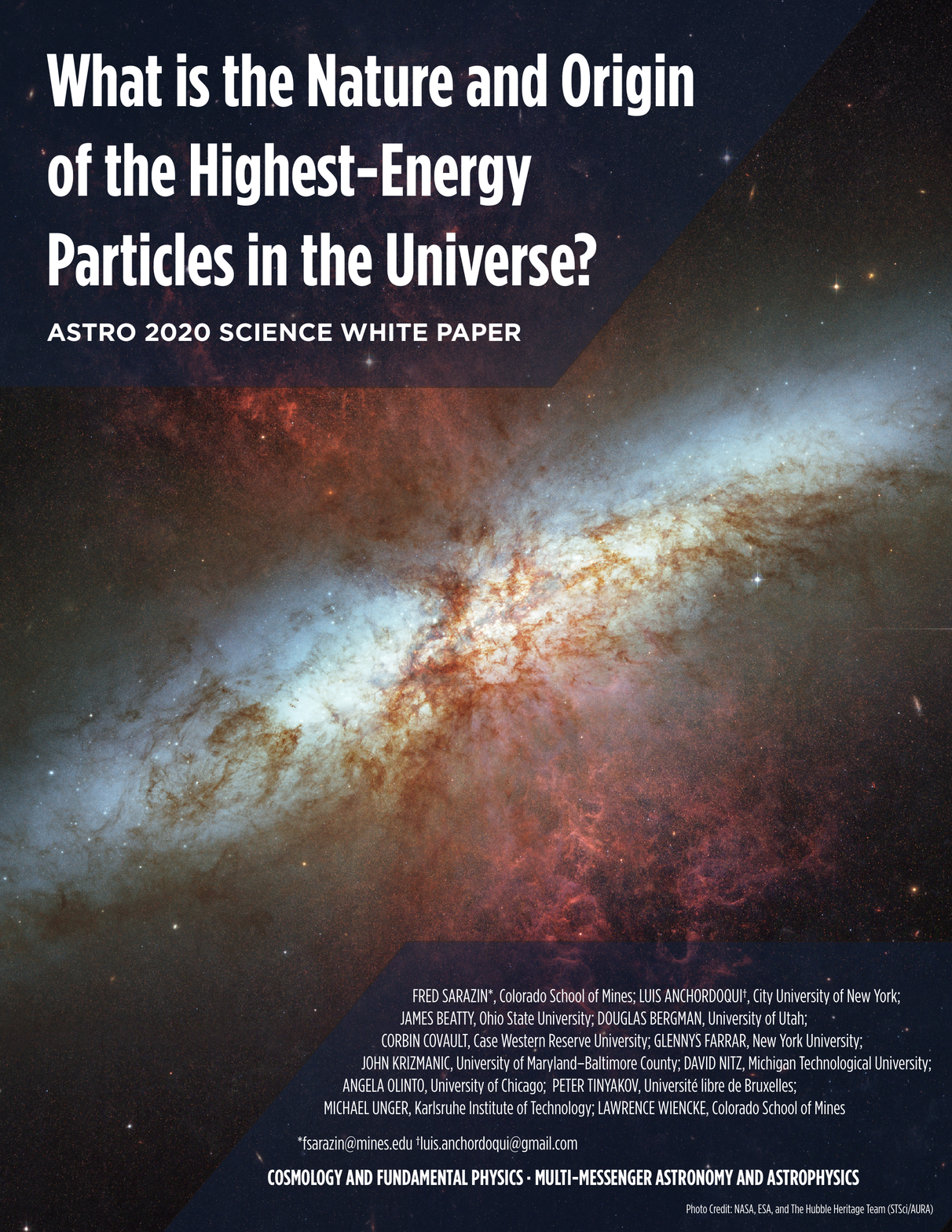}
\end{titlepage}

\pagebreak
\justifying


\section{The big questions and goals for the next decade}
\vskip-4mm
Ultra-High-Energy Cosmic-Ray (UHECR) astronomy in the next decade aims to answer the questions: {\it
    What is the nature and origin of UHECRs? 
    How are UHECRs accelerated to such extreme energies?
    Are there multiple types of sources and acceleration mechanisms?  
    Do UHECRs consist of both protons and heavier nuclei, and how does the composition evolve as a function of energy?}

In order to address these questions, the goals for the next decade will be to: identify one or more nearby UHECR sources, refine the spectrum and composition of the highest-energy Galactic and extragalactic cosmic-rays, exploit extensive air showers (EAS) to probe particle physics inaccessible at accelerators, and develop the techniques to make charged-particle astronomy a reality.

This agenda is largely achievable in the next decade, thanks to major experimental upgrades underway and new ground observatories and space missions in development. In combination with improved astrophysical neutrino statistics and resolution, the future UHECR observatories will enable a powerful multi-messenger approach to uncover and disentangle the common sources of UHECRs and neutrinos.   

\section{The UHECR paradigm shift}
\vskip-5mm
{\it {\bf Synopsis:} Results from the current large hybrid detectors have dispelled the pre-existing simple UHECR picture. A new paradigm is emerging and needs to be clarified and understood.}
\vskip2mm

The discoveries made over the past decade have transformed our understanding of UHECRs and their sources.  Prior to the development of very large, hybrid detectors, it was commonly believed that UHECRs were protons, that a spectral cutoff (if indeed there was one!) should be due to ``GZK'' energy losses on the cosmic microwave background~\cite{Greisen:1966jv,Zatsepin:1966jv}, that the Galactic-extragalactic transition would be marked by a shift from Galactic iron to extragalactic protons, and that the {\it ankle} feature at a few EeV marked the Galactic-extragalactic transition. We know now that this simple picture is mostly, if not entirely, wrong.

\begin{wrapfigure}{r}{0.47\textwidth}
    \centering
    \vspace{-9mm}
    \includegraphics[width=0.47\textwidth]{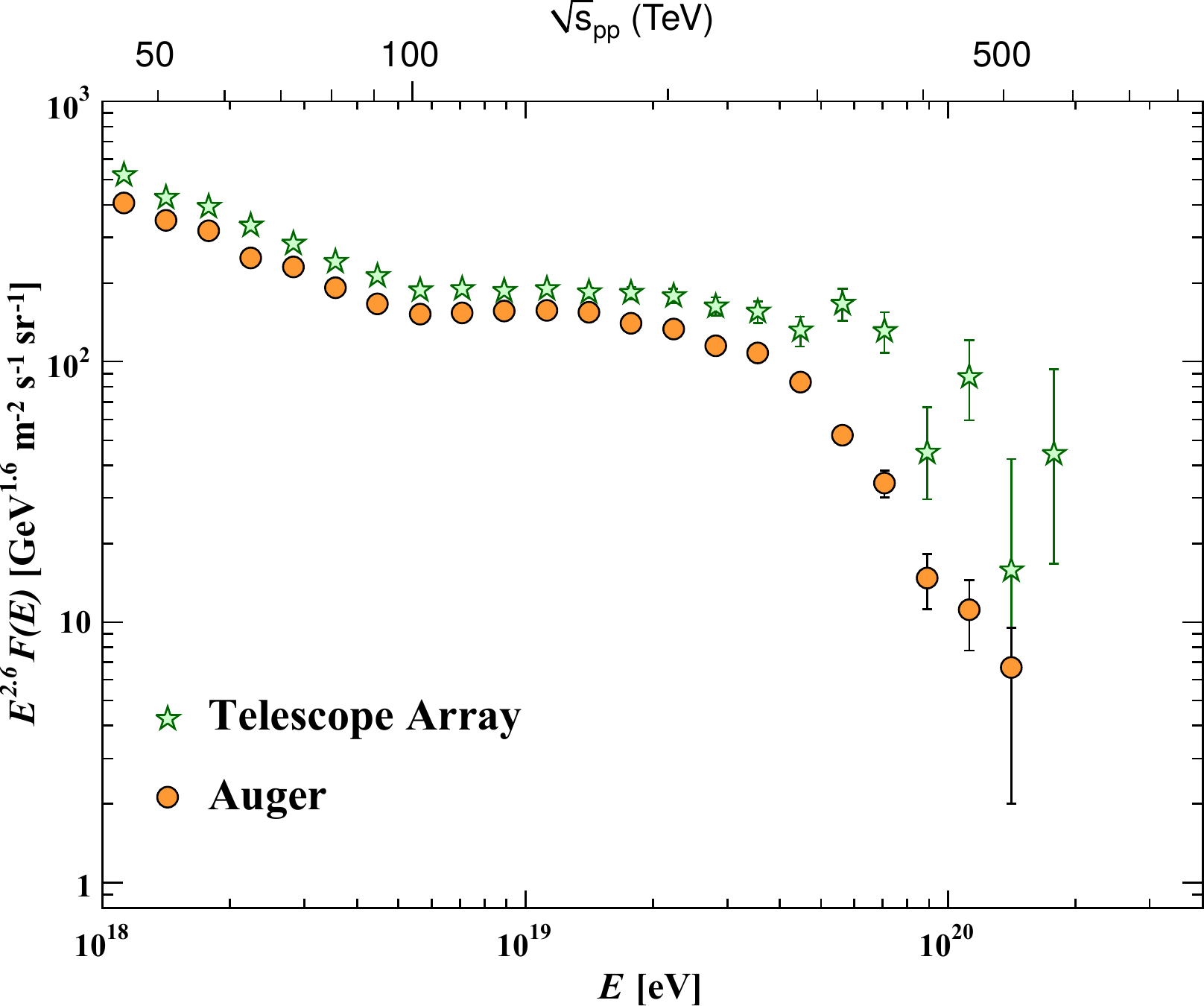}
    \vspace{-7mm}
    \caption{UHECR spectra measured 
    by TA and Auger, located in the northern and southern hemispheres, respectively (adapted from~\cite{Tanabashi:2018oca}).}
    \label{fig:spectrum}
    \vspace{-4mm}
\end{wrapfigure}
%
This revolution in our understanding was achieved thanks to: ({\it i})~hybrid detectors with air-fluorescence telescopes and surface detectors, plus improved measurement of fluorescence yields, giving much better energy calibration and providing greater sensitivity to composition; ({\it ii})~large aperture and high statistics, essential to reducing the systematics and particle-physics uncertainties in composition studies and providing sensitivity to tiny anisotropies in the UHECR arrival directions; and ({\it iii}) all-sky sensitivity thanks to detectors in both 
hemispheres, with overlapping regions of the sky.
Together, these advances enabled the spectrum, composition and anisotropies to be measured with higher resolution and smaller systematics from $10^{17}$ to above $10^{20}$~eV.

{\bf UHECR spectrum -- ankle and flux suppression: well established but not well explained}

As shown in Fig.~\ref{fig:spectrum}, the UHECR energy spectrum can be roughly described by a twice-broken power law~\cite{Abbasi:2007sv,Abraham:2008ru,Abraham:2010mj,AbuZayyad:2012ru,Aab:2017njo}. The first break is a hardening of the spectrum, known as ``the ankle.'' The second, an abrupt softening of the spectrum, may be interpreted as the long-sought GZK cutoff~\cite{Greisen:1966jv,Zatsepin:1966jv}, or else may correspond to the cosmic accelerators running out of steam  \cite{Allard:2008gj}. The differential energy spectra measured by the Telescope Array (TA) experiment and the Pierre Auger Observatory (Auger) agree within systematic errors below $10^{19}~{\rm eV}$. 
However, even after energy re-scaling, a large difference remains at and 
beyond the flux suppression~\cite{TheTelescopeArray:2018dje}. Once the significant differences in the common sky have been understood~~\cite{TheTelescopeArray:2018dje,Watson:2019clu}, fundamental differences between the northern and southern UHECR skies can be investigated~\cite{Abbasi:2018ygn,Abbasi:2017vru}.

{\bf UHECR primary composition  -- a more complex picture emerges}

The atmospheric column depth at which the longitudinal development of
a cosmic-ray shower reaches maximum, \Xmax, is a powerful observable
to determine the UHECR nuclear composition. Breaks in the elongation rate -- the rate of change of $\langle X_{\rm max} \rangle$ per decade of energy --  are associated to changes in the nuclear composition~\cite{Linsley:1981gh}, even when uncertainties in the UHE particle physics limit the accuracy of mapping between \Xmax\ and mass $A$.  The \Xmax{} measurements of both TA~\cite{Abbasi:2014sfa,Abbasi:2018nun}
and Auger~\cite{Abraham:2010yv,Aab:2014kda,Aab:2014aea,Aab:2017cgk}
indicate a predominantly light composition at around the ankle.  At the
highest energies (above 10 EeV), the Auger Collaboration reports a
significant decrease in the elongation rate, as well as a decrease of the shower-to-shower fluctuations of
\Xmax{} with energy.  Both effects suggest a gradual increase of the
average mass of cosmic rays with energy. Interpreting the
data with LHC-tuned hadronic interaction models gives a mean baryon
number $A \approx 14 - 20$ at $E \approx 10^{19.5}~{\rm eV}$. The Auger-TA
joint working group on composition concluded that the measurements of
the average shower maximum by TA and Auger are compatible within
experimental uncertainties at all
energies~\cite{Abbasi:2015xga,Hanlon:2018dhz}. The observed decrease 
of the standard deviation of the \Xmax~distributions reported
by Auger can currently neither be confirmed nor ruled out by TA
because of statistical limitations.   Thus the most recent data of 
UHE observatories reveals a complex evolution of the cosmic-ray composition with energy that challenges
the old simplistic models of CR sources.

{\bf UHECR Anisotropy -- where are the sources?}\\

\vskip-5mm
\begin{wrapfigure}{r}{0.53\textwidth}
    \vskip-8.0mm
    \centering
 \includegraphics[width=0.53\textwidth]{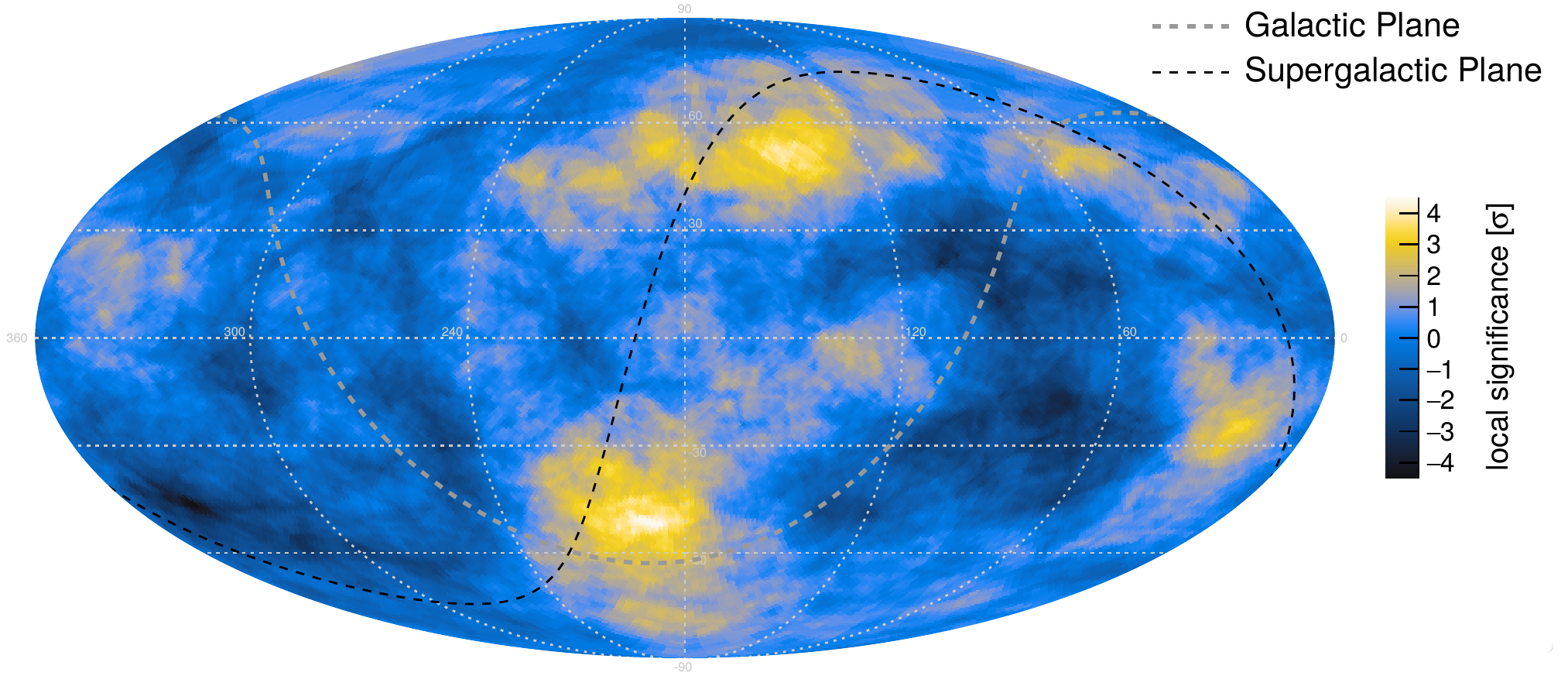}
    \vspace{-3.0mm}
    \caption{Sky map, in equatorial coordinates, of local over- and under-densities in units of standard deviations of UHECRs above $47\pm 7~{\rm EeV}$. Taken from~\cite{Biteau}. }
    \label{fig:skymap}
    \vspace{-4.4mm}
\end{wrapfigure}
Composition measurements have led to a paradigm shift, with cosmic rays now
understood to be light (proton dominated) near $10^{18}$~eV and evolving towards heavier composition with increasing energy, spanning a narrow range of atomic masses at each energy.  Below the ankle, the arrival directions are highly isotropic~\cite{ThePierreAuger:2014nja}, arguing that these protons must be of extragalactic origin. They are consistent with being secondary products of the photo-disintegration of UHECR
nuclei in the environment of their sources~\cite{Unger:2015laa} and/or can share the origin of PeV neutrinos~\cite{Fang:2017zjf,Kachelriess:2017tvs}.  At higher energies,  Galactic and extragalactic deflections of UHECR nuclei are expected to smear point sources into warm/hot spots, for which evidence is accumulating. TA has recorded an excess above the isotropic background-only expectation in cosmic rays with energies above $10^{19.75}~{\rm eV}$~\cite{Abbasi:2014lda,Kawata:2015whq}, while 
Auger has reported a possible correlation with nearby starburst galaxies, with a (post-trial) $4\sigma$ significance, for events above $10^{19.6}~{\rm eV}$~\cite{Aab:2018chp}. A slightly weaker association ($2.7 \sigma$) with active galactic nuclei emitting $\gamma$-rays is also found in Auger events above $10^{19.78}~{\rm eV}$~\cite{Aab:2018chp}.  A blind search for anisotropies combining Auger and TA data has been recently carried out, with the energy scales equalized by the flux in the common declination 
band~\cite{Biteau}. The most-significant excess is obtained for a $20^\circ$ search radius, with a global (post-trial) significance of $2.2\sigma$. The local (Li-Ma~\cite{Li:1983fv}) significance map of this study is shown in Fig.~\ref{fig:skymap}. The tantalizing visual correlation of 
high-significance regions with the supergalactic plane is currently under study within the Auger/TA anisotropy working group.

{\bf UHECR $\leftrightharpoons$ neutrino -- the missing GZK neutrinos and targets of opportunity} 

The non-observation of neutrino
candidates beyond background expectations above $10^{16}$~eV  by IceCube~\cite{Aartsen:2018vtx},
Auger~\cite{Aab:2015kma}, and ANITA~\cite{Gorham:2019guw}  severely
constrains the magnitude of the very high-energy neutrino flux. This flux
has a nearly guaranteed component from the decays of pions produced by UHECR protons 
interacting en route to Earth~\cite{Beresinsky:1969qj}. The accumulation
of these neutrinos over cosmological time,  known as the cosmogenic
neutrino flux,  constitutes a powerful
tool of the multi-messenger program. 
IceCube, Auger, and ANITA limits already challenge models in which the highest-energy UHECRs
are proton-dominated~\cite{Ahlers:2009rf,Heinze:2015hhp,Aartsen:2016ngq,Moller:2018isk,AlvesBatista:2018zui,vanVliet:2019nse,Heinze:2019jou,Berezinsky:2016jys}. Additionally, UHECR experiments add the capability of searching for neutrinos from target-of-opportunity 
events~\cite{Aab:2016ras,GBM:2017lvd,ANTARES:2017bia}. 

{\bf Closing the loop -- Particle physics with UHECRs}

Essential to accurate composition determination is the correct understanding of the physics of EASs, which requires accurate modeling of particle physics at center-of-mass energies up to hundreds of TeV -- far beyond the 14~TeV reach of the LHC.  Internal-consistency studies of EASs show that state-of-the-art LHC-tuned hadronic event generators do not correctly reproduce in detail the multitude of observables that can be probed by UHECR detectors~\cite{Aab:2016hkv,Abbasi:2018fkz}. 
Upgraded and next-generation experiments  
are designed to extend our understanding of hadronic interactions well into the hundreds of TeV regime~\cite{Cazon:2018gww,Cazon:2018bvs,Dembinski:2019uta,Baur:2019cpv}.  This will increase the accuracy in determining the UHECR composition, and be a boon to particle physics. The column energy-density in UHECR-air collisions is an order of magnitude greater than in Pb-Pb collisions at the LHC~\cite{Farrar:2019cid}, suggesting the potential for new hadronic physics from gluon saturation and the possibility of exploring quark-gluon plasma (QGP) at far higher energies than available in accelerators~\cite{Farrar:2013sfa,Anchordoqui:2016oxy,ALICE:2017jyt}. To find out more about the latest results on UHECRs, see e.g.~\cite{Kotera:2011cp,Anchordoqui:2018qom, miappReview2019}.

\section{Identifying candidate sources for extreme accelerators}
\vskip-5mm
{\it {\bf Synopsis:} The high-energy (HE) astrophysics community remains abreast with the evolving observational picture and has developed a wide variety of new exciting models that will be further tested by the data collected over the next decade.}
\vskip2mm
An even greater diversity of sources and acceleration mechanisms is now under consideration as a result of theory advances and the evolving observational picture.  If the highest-energy UHECRs are exclusively intermediate mass nuclei, as is consistent with present data, the demands on accelerators are considerably eased compared to a pure-proton scenario, because the maximum required rigidity $R=E/Z$ and the bolometric luminosity required of the candidate sources are reduced; here $Z$ is the charge of the UHECR in units of the proton charge. 
Further refinements in measuring the composition evolution and possible composition anisotropy are crucial to source inference.

Rapid progress in computational HE astrophysics is dramatically advancing the study of acceleration mechanisms.  Some of the current contenders for acceleration mechanisms and source types are: shock acceleration~\cite{Krymskii:1977,Axford:1977,Bell:1978zc,Bell:1978fj,Blandford:1978ky,Lagage:1983zz,Drury:1983zz,Blandford:1987pw}, in systems ranging from the large scale shocks surrounding galaxy clusters~\cite{Norman:1995,Kang:1996rp,Ryu:2003cd} to internal or external shocks of starburst-superwinds~\cite{Anchordoqui:1999cu,Anchordoqui:2018vji}, AGN~\cite{Biermann:1987ep,Takahara:1990he,Rachen:1992pg,Romero:1995tn,GopalKrishna:2010wp,Blandford:2018iot,Matthews:2018laz,Matthews:2018rpe,Matthews:2019bkp} or GRB~\cite{Waxman:1995vg,Vietri:1995hs,Dermer:2006bb,Wang:2007xj,Murase:2008mr,Baerwald:2013pu, Globus:2014fka,Zhang:2017moz} jets, and the jets of tidal disruption events (the transient cousins of AGN jets)~\cite{Farrar:2008ex,Farrar:2014yla,Pfeffer:2015idq}.   Other contenders are: shear acceleration~\cite{Rieger:2004jz,Kimura:2017ubz} and one-shot mechanisms such as ``espresso"~\cite{Caprioli:2015zka}  in which an AGN or other jet boosts a galactic CR of the host galaxy;
    EMF acceleration as in fast-spinning pulsars~\cite{Blasi:2000xm,Fang:2012rx,Fang:2013cba} and magnetars~\cite{Arons:2002yj}, black holes~\cite{Blandford:1977ds,Znajek,Neronov:2007mh}, and potentially reconnection, explosive reconnection, gap and/or wakefield acceleration~\cite{Chen:2002nd,Murase:2009pg,Ebisuzaki:2013lya}.

The multitude of possibilities suggests there may well be multiple sources of UHECRs, some of which may be transient, making the identification of sources even more challenging and essential. 
Anticipating the advent of UHECR-astronomy thanks to composition-tagging and better understanding of the Galactic magnetic field, we can foresee having access to the UHECR spectrum of individual sources. 
That will be key to determining the acceleration mechanism(s) and identifying the potential sources, whether those are steady or transient~\cite{note_1}, much as spectra at X-ray and $\gamma$-ray wavelengths have clarified the workings of blazars and their kin. 

\section{Stepping up to the new challenges} 
\vskip-5mm
{\it {\bf Synopsis:} Future discoveries will be made through a combination of enhanced statistics, refined analyses afforded by upgraded observatories and next-generation experiments, and the additional constraints provided by multi-messenger astrophysics.}
\vskip2mm
The more complex picture that has emerged over the past decade presents a challenge to discovering UHECR sources and unraveling how UHECRs are accelerated -- the holy grail of multi-messenger astrophysics for decades.  
Yet, after about a decade of operation, both Auger and TA have provided tantalizing evidence that new discoveries are within reach. In this context, the discovery of a large-scale asymmetry in the arrival direction 
distribution of events recorded by the Auger~\cite{Aab:2017tyv,Aab:2018mmi} (statistical significance $>5\sigma$) represents a compelling example of the power of accumulating more statistics.  By 2025, Auger will roughly double the size of the sample for which the 4$\sigma$ correlation with starburst galaxies was observed~\cite{Aab:2018chp}, allowing for an independent test of the starburst hypothesis. Combining the data samples Auger may actually reach a statistical significance $> 5\sigma$ by 2025. For TA, a significant increase of exposure will allow the northern hemisphere hot spot to be adequately explored.

{\bf The path to new discoveries -- increased exposure and higher sensitivity}

Both Auger and TA are undergoing upgrades to respond to the evolving observational picture. TA$\times$4 is designed to cover the equivalent of Auger's aperture~\cite{Kido}, to allow for a 5$\sigma$ 
observation of the northern hot spot by 2026 or so. Auger's upgrade (``AugerPrime''~\cite{Aab:2016vlz}) 
focuses on more detailed measurements of each shower observed.
This will enable event-by-event probabilistic composition assignment (hence selection of low-$Z$ events), enhanced capacity to study UHE hadronic interactions, and increased sensitivity to high-energy neutrinos~\cite{Aab:2016eeq,Aab:2018ytv}. Both upgrades contribute to an overall strategy comprising  three broad approaches:\\ 
$\bullet$  {\it Detailed information on the composition of UHECRs as a function of energy} can eliminate some source candidates from contention as a dominant contributor, while combining the UHECR spectrum and composition with neutrino and $\gamma$-ray spectra, produces powerful constraints on the environment surrounding the sources.  \\
$\bullet$   {\it Composition-assisted anisotropy studies add new potential to identify individual UHECR sources.} With a mixed composition, UHECRs can experience large deflections by Galactic and extragalactic magnetic fields, even at the highest energies.  Composition-tagging allows the subset of events with highest rigidity and hence smallest deflections to be selected, to strongly enhance source identification both with UHECRs alone and in combination with neutral messengers.\\
$\bullet$ {\it Neutral-messenger arrival direction correlations} can identify individual sources, 
while their temporal associations are sensitive to flaring sources.
However, correlations between $\nu$'s and $\gamma$'s alone cannot give a complete picture.  Even if some blazars produce UHECRs, the majority of the AGN jets 
are beamed away from Earth, 
thus not all sources may be observable with neutral particles.

\begin{wrapfigure}{r}{0.52\textwidth}
    \vskip-6mm
    \centering
    \includegraphics[width=0.50\textwidth]{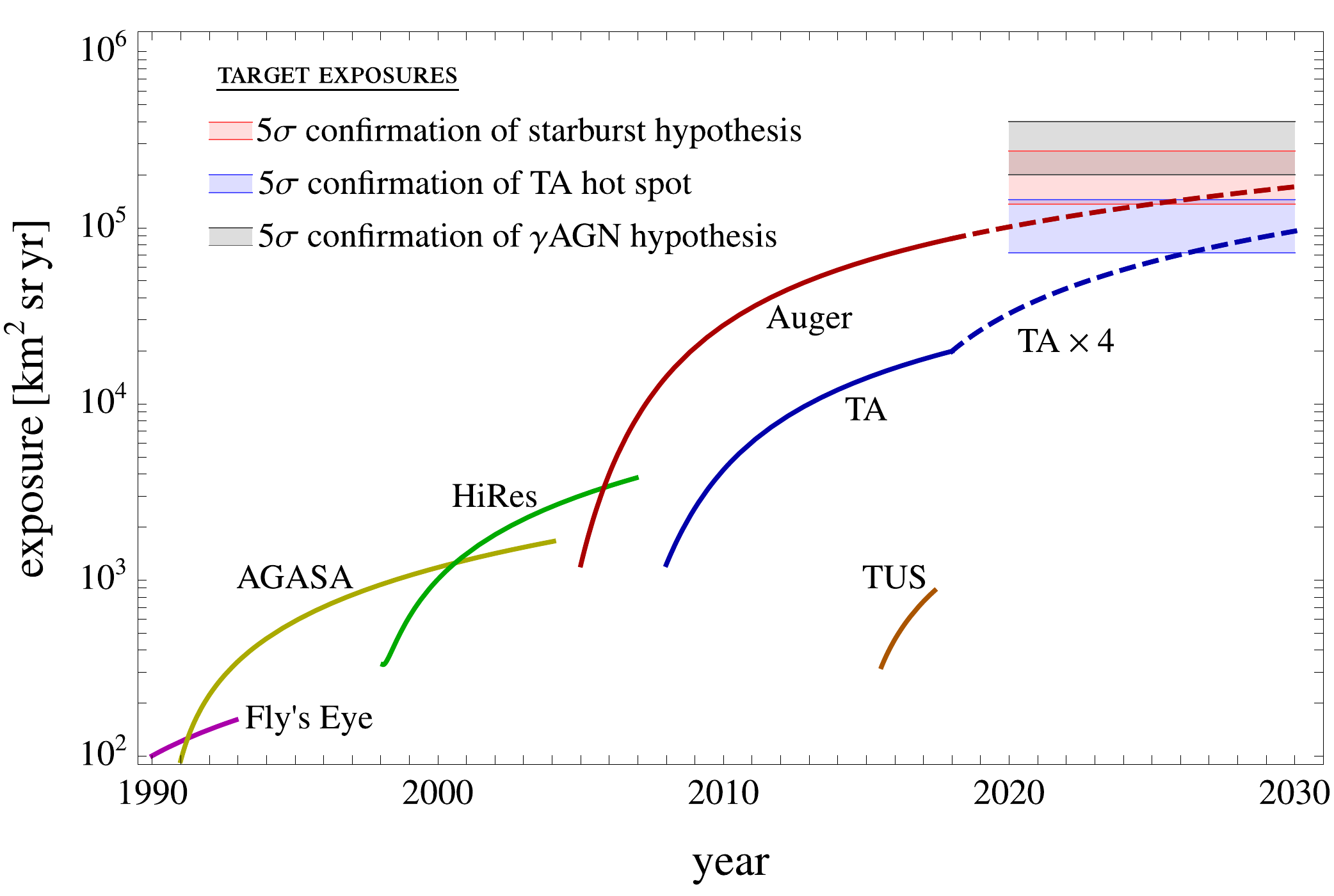}
    \vspace{-3mm}
    \caption{The historical growth of UHECR integrated exposures~\cite{Bird:1994wp,Takeda:2002at,AbuZayyad:2002sf,Khrenov:2017eev,AbuZayyad:2012ru,Aab:2017njo}. The projected integrated exposures of Auger and TA$\times$4 are shown together with the target exposure bands that could achieve a 5$\sigma$ observation of the northern hot spot (TA/blue) and $5\sigma$ confirmation of the starburst and $\gamma$AGN hypotheses (Auger/red and Auger/black). The lower edge of the bands indicates the required exposure of an experiment which permanently observes the same region of the sky as Auger/TA, whereas the upper edge indicates the required exposure for a full-sky instrument which observes the Auger/TA sky only half of the time. Figure by J.F.~Soriano (CUNY).}
    \label{fig:exposure}
    \vspace{-4mm}
\end{wrapfigure}

By studying the time evolution of the test statistics of the anisotropy searches~\cite{Abbasi:2014lda,Kawata:2015whq,Aab:2018chp}, we have projected the required exposures to obtain a $5\sigma$ confirmation of the various hypotheses. These target exposures, shown in Fig.~\ref{fig:exposure}, demonstrate that  the continued operation into the next decade of the upgraded observatories will bring us within reach of new discoveries. Based on the existing data, new target exposures and new resolutions on key observables can be inferred and should become the basis for the design of next-generation 
ground observatories and space instruments. 

{\bf Requirements to achieve the science goals and next generation UHECR experiments}
~~~Complementing the upgraded Auger and TA detectors, the next generation of UHECR instruments focusing on the flux suppression region ($E \gtrsim 10^{19.6}~{\rm eV}$) will need to achieve:
{\it (i)}~Significant gain in exposure, from Fig.~\ref{fig:exposure} we estimate $\sim 5 \times 10^5~{\rm km}^2 \,  {\rm sr} \, {\rm yr}$ to allow for a $5\sigma$ observation of all potential signals.
{\it (ii)}~A resolution $\Delta X_{\rm max} \sim 20~{\rm g/cm^2}$~\cite{note_2}. (Note that $\langle X_{\rm max} \rangle$ of $p$ and Fe are separated by $\sim 100~{\rm g/cm^2}$~\cite{Kampert:2012mx}, thus the recommended $\Delta X_{\rm max}$ would allow studies with a four-component nuclear composition model~\cite{Krizmanic:2013tea,Krizmanic:2015icrc,Soriano:2018lly,note_3}.) {\it (iii)}~An energy resolution ideally $\Delta E/E \lesssim 20\%$  to limit the effects of lower-energy event spillover near the flux suppression~\cite{Brummel:2013icrc}. {\it (iv)}~An angular resolution comparable to that of previous experiments to both test the hints for intermediate-scale anisotropies in Auger and TA data and continue the search for small-scale clustering. {\it (v)}~Full sky coverage to
 test the hints of  declination dependence of the TA
spectrum~\cite{Abbasi:2018ygn,Abbasi:2017vru}. 

At present, the most advanced concept in pursuit of these objectives is the Probe of Extreme Multi-Messenger Astrophysics (POEMMA) satellites~\cite{Olinto:2017xbi}.
 POEMMA will reach $\sim 2.5 \times 10^5$ km$^2$ sr yr exposure in 5 years with (calorimetric) stereo EAS reconstruction that significantly improves the angular, energy, and \Xmax\ resolutions over that from monocular space-based EAS measurement. 
 A factor $\times$10 increase over current UHECR apertures may be achievable on the ground; e.g., the GRAND project  
 has been designed to use low-cost radio antennas deployed over 200,000 km$^{2}$ to measure highly-inclined EASs from UHE cosmic-rays and neutrinos~\cite{Alvarez-Muniz:2018bhp}.

In the new era of multi-messenger astronomy, improved measurements of the highest-energy particles will  
provide a compelling and complementary view of the extreme universe.  The UHECR community is aggressively responding to the new questions posed by the UHECR paradigm shift.  The next decade will test the hints of source candidates and build the next-generation experiments that will usher in a new era of charged-particle astronomy.

\pagebreak

\pagebreak
{\Large \bf Endorsements:}

\vskip5mm

212 scientists from 27 countries provided their support to this white paper (including 10 after the deadline). The complete list of endorsers is below:

\vskip5mm

Tareq AbuZayyad, James Adams, Markus Ahlers, Roberto Aloisio, Jaime Alvarez-Muniz, Rafael Alves Batista, Karen Andeen, Sofia Andringa, Ignatios Antoniadis, Carla Aramo, Hernan Asorey, Pedro Assis, Reda Attallah, Xinhua Bai, Vernon Barger, Sebastian Baur, Mario Edoardo Bertaina, Peter Bertone, Dave Besson, Francesca Bisconti, Jonathan Biteau, Jiri Blazek, Martina Bohacova, Carla Bonifazi, Olga Botner, Antonio Bueno, Mauricio Bustamante, Karen Salomé Caballero-Mora, Damiano Caprioli, Juan Miguel Carceller, Rossella Caruso, Marco Casolino, Antonella Castellina, Lorenzo Cazon, Yaocheng Chen, Koun Choi, Eugene Chudnovsky, Roger Clay, Alan Coleman, Ruben Conceição, Stephane Coutu, Bruce Dawson, Sijbrand De Jong, João de Mello Neto, Ivan De Mitri, Vitor de Souza, Cosmin Deaconu, Valentin Decoene, Luis Del Peral, Hans Dembinski, Peter B. Denton, Tyce DeYoung, Rebecca Diesing, Carola Dobrigkeit, Michael DuVernois, Toshikazu Ebisuzaki, Rikard Enberg, Ralph Engel, Johannes Eser, Alberto Etchegoyen, Jonathan L. Feng, Jorge Fernandez Soriano, Brian Fick, Gustavo Figueiredo, George Filippatos, Christer Fuglesang, Toshihiro Fujii, Thomas Gaisser, Beatriz Garcia, Carlos Garcia Canal, Noemie Globus, Jonas Glombitza, Nicolas Gonzalez, Darren Grant, Fausto Guarino,  Allan Hallgren, Robert Halliday, Francis Halzen, Diego Harari, John Harton, Andreas Haungs, Dan Hooper, Tim Huege, Naoya Inoue, Susumu Inoue, Dmitri Ivanov, Jeffrey Johnsen, Eleanor Judd, Jakub Jurysek, Daniel Kabat, Michael Kachelriess, Fumiyoshi Kajino, Toshitaka Kajino, Marc Kamionkowski, Karl-Heinz Kampert, Donghwa Kang, Teppei Katori, Bianca Keilhauer, Azadeh Keivani, Matthias Kleifges, Kumiko Kotera, Ely Kovetz, Viktoria Kungel, Evgeny Kuznetsov, Mikhail Kuznetsov, John Learned, Simon Mackovjak, Max Malacari, Paul Mantsch, Danny Marfatia, Ioana Maris, Giovanni Marsella, Oscar Martinez-Bravo, James Matthews, Eric Mayotte, Peter Mazur, Gustavo Medina Tanco, Kevin-Druis Merenda, Jamal Mimouni, Lino Miramonti, Miguel Mostafa, Kohta Murase, Ryan Nichol, Dalibor Nosek, Matthew J. O'Dowd, Foteini Oikonomou, Giuseppe Osteria, A. Nepomuk Otte, Sergio Palomares-Ruiz, Mikhail Panasyuk, Etienne Parizot, Francesco Perfetto, Mário Pimenta, Tsvi Piran, Raul R. Prado, Paolo Privitera, Maxim Pshirkov, Sean Quinn, Julian Rautenberg, Patrick Reardon, Mary Hall Reno, Marco Ricci, Felix Riehn, Markus Risse, Vincenzo Rizi, Maria Dolores Rodriguez Frias, Markus Roth, Grigory Rubtsov, Naoto Sakaki, Takashi Sako, Marcos Santander, Eva Santos, Christian Sarmiento-Cano, Michael Schimp, David Schmidt, Olaf Scholten, Frank Schroeder, Sonja Schroeder, Valentina Scotti, David Seckel, Dmitri Semikoz, Ronald Cintra Shellard, Kenji Shinozaki, Ogio Shoichi, Günter Sigl, Lorenzo Sironi, Radomir Smida, Dennis Soldin, Paul Sommers, David Spergel, Glenn Spiczak, Jaroslaw Stasielak, Floyd Stecker, Andrew Strong, Alberto Daniel Supanitsky, Alvaro Taboada, Yoshiyuki Takizawa, Alex Tapia, Andrew Taylor, Charles Timmermans, Carlos José Todero Peixoto, Diego F. Torres, Delia Tosi, Petr Travnicek, Yoshiki Tsunesada, Sara Turriziani, Ralf Ulrich, Jose Valdes-Galicia, Inés Valiño, Piero Vallania, Justin Vandenbroucke, Arjen van Vliet, Darko Veberic, Tonia Venters, Jakub Vicha, Abigail Vieregg, Alex Vilenkin, Serguei Vorobiov, Alan Watson, Eli Waxman, Henryk Wilczynski, Martin Will, Walter Winter, Stephanie Wissel, Brian Wundheiler, Tokonatsu Yamamoto, Alexey Yushkov, Danilo Zavrtanik, Lukas Zehrer, Arnulfo Zepeda, Jianli Zhang, Mikhail Zotov


\begin{thebibliography}{99}
\bibitem{note_2} It is important to keep in mind that
   the {\it intrinsic} $X_{\rm max}$ fluctuations are
roughly $20~{\rm g/cm^2}$ for Fe growing to $60~{\rm g/cm^2}$ for $p$. Consequently, a $\Delta X_{\rm max} < 20~{\rm g/cm^2}$ will not improve the event-by-event composition 
capabilities. Once the level for intrinsic fluctuations has been reached, the merit factor can only be increased by adding another observable (e.g. muons).
\end{thebibliography}
\end{document}